\documentclass[11pt]{article}
\usepackage{graphics,cite,amssymb,epsfig,float}
\usepackage[usenames,dvips]{color}
\usepackage{amsmath}

\usepackage{multirow}
\makeatletter

\@addtoreset{equation}{section}
\makeatother

\def\lsim{\;\raise0.3ex\hbox{$<$\kern-0.75em\raise-1.1ex\hbox{$\sim$}}\;}
\def\gsim{\;\raise0.3ex\hbox{$>$\kern-0.75em\raise-1.1ex\hbox{$\sim$}}\;}

\def\ben{\begin{enumerate}}  \def\een{\end{enumerate}}
\def\bit{\begin{itemize}}    \def\eit{\end{itemize}}
\def\beq{\begin{equation}}   \def\eeq{\end{equation}}
\def\ba{\begin{array}}       \def\ea{\end{array}}
\def\bea{\begin{eqnarray}}   \def\eea{\end{eqnarray}}

\newcommand{\comment}[1]{}

\begin{document}

\begin{titlepage}
\renewcommand{\thefootnote}{\fnsymbol{footnote}}
\setcounter{footnote}{0}

\vspace*{-2cm}
\begin{flushright}
LPT Orsay 11-60 \\

\vspace*{2mm}
\today
\end{flushright}

\vspace*{2mm}

\begin{center}
\vspace*{15mm}
{\Large\bf Enhanced Higgs Mediated Lepton Flavour Violating Processes
in the Supersymmetric Inverse Seesaw Model} \\
\vspace{1cm}
{\bf Asmaa Abada, Debottam Das and 
 C\'{e}dric Weiland
}

 \vspace*{.5cm} 
 Laboratoire de Physique Th\'eorique, CNRS -- UMR 8627, \\
Universit\'e Paris-Sud 11, F-91405 Orsay Cedex, France
\end{center}

\vspace*{10mm}
\begin{abstract}

\vspace*{3mm}
We study the impact of the inverse seesaw mechanism on several low-energy 
flavour 
violating observables such as $\tau\rightarrow \mu\mu\mu$  in the context of the 
Minimal Supersymmetric Standard Model. As a consequence of the 
inverse seesaw, the contributions of the right-handed 
sneutrinos significantly enhance the Higgs-mediated
penguin diagrams. We find that  
different flavour violating 
branching ratios can be enhanced by as much as two orders of magnitude.
We also comment on the impact of the Higgs-mediated processes on the
leptonic $B$-meson decays and  on the Higgs flavour violating decays. 

\end{abstract}

\vspace*{3mm}

\renewcommand{\thefootnote}{\arabic{footnote}}
\setcounter{footnote}{0}

\vspace*{5mm}
\end{titlepage}\setcounter{page}{2}
\section{Introduction}
Neutrino oscillations have provided indisputable evidence for flavour violation in the neutral
lepton sector. In the absence of any fundamental principle that prevents charged lepton flavour 
violation, one expects that extensions of the Standard Model (SM) accommodating neutrino masses and
mixings should also allow for lepton flavour violation (LFV) in the charged lepton sector. 
Indeed, the additional new flavour dynamics and new field content present in many extensions of the 
SM may provide contributions to charged LFV (cLFV) processes such as radiative (e.g. $\mu\to e\gamma$) 
and three-body lepton decays (for instance $\tau\to \mu\mu\mu$). 
These decays generally arise from higher order processes,
and so their branching ratios (Brs) are expected to be small, making them difficult to observe. Thus,  any cLFV 
signal would provide clear evidence for new physics: mixings in the lepton sector and probably the presence of new particles, 
possibly shedding light  on  the origin of neutrino mass generation.

The search for manifestations of charged LFV constitutes the goal of several experiments 
~\cite{Bellgardt:1987du, Brooks:1999pu, Aysto:2001zs, 
Aubert:2003pc, Akeroyd:2004mj, Aubert:2005ye, Aubert:2005wa, 
Kuno:2005mm, Ritt:2006cg, Bona:2007qt, Hayasaka:2007vc, Kiselev:2009zz, PRIME
}, 
 dedicated to look for signals of processes such as rare
radiative decays,  three-body decays and muon-electron conversion in nuclei. 
Despite the fact that a cLFV signal could provide  clear evidence for 
new physics, the underlying mechanism of lepton mixing might be difficult to unravel. 
In  parallel to the low-energy searches for new physics, i.e. via indirect effects of possible new particles,  
the LHC has started to search directly for these new particles in its quest to unveil the 
mechanism of electroweak symmetry breaking, thus possibly providing a solution to the SM hierarchy problem.

Among the many possible extensions of the SM, supersymmetry (SUSY)  is a well motivated solution for the 
hierarchy problem,  providing many other appealing aspects such as gauge coupling unification and 
dark matter candidates.  If the LHC experiments indeed discover SUSY, it is then
extremely interesting to consider supersymmetric models that can also explain neutrino masses and
mixings. Furthermore, it is only natural to expect that such models  might also give rise to cLFV.
If SUSY is indeed realised in nature, cLFV (mediated by new sparticles) would  provide a 
new probe to explore the origin of lepton mixings, playing a complementary 
r\^ole to other searches 
of new physics, i.e. LHC direct searches and neutrino dedicated experiments.

One of the most economical possibilities is  to embed a seesaw 
mechanism~\cite{seesaw:I, seesaw:II, seesaw:III} 
in the framework of SUSY models (i.e.~the SUSY seesaw) \cite{susy-seesaw}. For any seesaw realisation, the
neutrino Yukawa couplings could leave their imprints in the  SUSY soft-breaking slepton mass matrices, and 
consequently induce flavour violation at low energies due to the renormalisation group (RG)
evolution of the SUSY soft-breaking parameters. 
The caveat of these scenarios is that, in
order to have sufficiently large Yukawa couplings (as required to account for large cLFV Brs), 
the typical scale of the extra particles
(such as  right handed neutrinos, scalar or fermionic isospin triplets) is in general very high, potentially very close to the gauge coupling unification scale.   
However, such a high (seesaw) scale would be impossible to probe experimentally.

On the other hand,  the so-called inverse seesaw \cite{inverse} constitutes a very appealing alternative to the 
"standard" seesaw realisations.
Embedding an inverse seesaw mechanism in the Minimal Supersymmetric extension of the SM (MSSM) requires the inclusion of 
two additional gauge singlet superfields, with opposite lepton numbers ($+1$ and $-1$). 
When compared to other SUSY seesaw realisations, cLFV observables are 
enhanced in this framework , 
and such an enhancement can be attributed to 
large neutrino Yukawa couplings  ($Y_\nu \sim O(1)$), compatible with a seesaw 
scale $M$, close to the  
electroweak one, thus within LHC reach. 

The  differences between  the inverse seesaw and  the standard one can be conceptually 
understood from an effective point of view and linked to 
the distinct properties of the 
lepton number {\em violating} dimension-5 (Weinberg) operator (responsible for neutrino masses and mixings) 
and the total lepton number {\em conserving} dimension-6 operator, which is at the origin of cLFV. 
Contrary to what occurs in the standard seesaw, these two operators are
de-correlated in the inverse seesaw, implying that the suppression of the coefficient of the 
dimension-5 operator will not affect the size of the coefficient of the dimension-6 operator. 
In both seesaws, the latter operator is proportional to  
$\left(Y_\nu^{\dagger}\frac{1}{\left|M\right|^{2}}Y_\nu\right)$;   
however, in the case of a type I seesaw, the dimension-5 operator is proportional to 
$\left(Y_{\nu}^{\dagger}\frac{1}{M}Y_{\nu}\right)$, while in the case of an inverse seesaw, it has a further suppression of $ \frac{\mu} {M}$ ($\mu$ being a dimensionful parameter, linked to the mass of the sterile singlets). 
The dimension-6 operator  will thus be extremely suppressed in the case of a type I seesaw, since in this case  $M$ is very 
large to accommodate natural $Y_\nu$. In contrast,  in the inverse seesaw, small neutrino masses 
can easily be accommodated via tiny values of $\mu$, which will not  affect the dimension 6 operator.
Furthermore, such small values of $\mu$ are natural in the sense of 't~Hooft since in the limit where 
$\mu\to 0$, the total lepton number symmetry is restored~\cite{tHooft}.

In view of the strong potential of the inverse seesaw mechanism  regarding cLFV, 
several phenomenological studies have recently been carried out~\cite{Deppisch:2004fa,Deppisch:2005zm,Garayoa:2006xs,Arina:2008bb,
Ma:2009gu,Dev:2009aw,Malinsky:2009df,Bazzocchi:2009kc,Hirsch:2009ra,
Bazzocchi:2010dt}.
While a non-supersymmetric inverse seesaw requires two pairs of singlets to explain neutrino oscillation data~\cite{Malinsky:2009df},  the supersymmetric
generalization can accommodate neutrino data~\cite{Hirsch:2009ra} with
just one pair of singlets. The latter scenario is also known as 
the minimal supersymmetric inverse seesaw model (MSISM). This model can 
also comply with the dark matter relic abundance of the Universe~\cite{Arina:2008bb}. 

The extra 
TeV scale singlet neutrinos may significantly contribute to cLFV observables, 
irrespective of the supersymmetric states \cite{valle_heavy_lepton}. 
Supersymmetric realisations of the inverse seesaw may enhance these cLFV rates 
even further 
(e.g. the contributions to $l_i \rightarrow l_j \gamma$, which 
has been analysed in \cite{Deppisch:2004fa}). 
Furthermore, this seesaw model can 
have  LHC signatures: the extra singlets can participate in the decay chains,  
 leading to effects which can be important, particularly in the case in which one of the singlets is the lightest
supersymmetric particle (LSP)~\cite{Hirsch:2009ra}.

In this paper, we focus on contributions to cLFV observables,  such as $\tau\to \mu\mu\mu$, arising from a 
Higgs-mediated  effective vertex.  We explore the contributions which are due to the presence of comparatively light right-handed 
neutrinos and sneutrinos (which are usually negligible in the framework of a type I SUSY-seesaw), while still having large 
neutrino Yukawa couplings. We find that all these contributions can lead to a significant enhancement of 
several cLFV observables.

The paper is organised as follows. 
In Section~\ref{sec:mod}, we define the model, 
providing a brief overview on the implementation of the 
inverse seesaw in the 
MSSM. In Section~\ref{lfv:diag}, we  discuss the implications of this model regarding 
low-energy cLFV observables, particular emphasis being given to  the Higgs-mediated
processes. 
In Section~\ref{hmlfv}, we study the Higgs-mediated  contributions to several lepton flavour violating observables 
and compare our results to  present bounds and to  future experimental sensitivities in Section~\ref{sec:discussion}. 
Then we draw some generic conclusions on the viability of an inverse seesaw as the
underlying mechanism of LFV. We finally conclude in Section~\ref{sec:conclusions}.

\section{Inverse Seesaw Mechanism in the MSSM}\label{sec:mod}

The model consists of the MSSM extended 
by three pairs of singlet superfields, $\widehat{\nu}^c_i$ and $\widehat{X}_i$ ($i=1,2,3$)\footnote{ $\widetilde{\nu}^c=\widetilde{\nu}_R^* $}
with lepton numbers assigned to be $-1$ and $+1$, respectively. 
The supersymmetric inverse seesaw model is defined by the superpotential

\bea
{\mathcal W}&=& \varepsilon_{ab} \left[
Y^{ij}_d \widehat{D}_i \widehat{Q}_j^b  \widehat{H}_d^a
              +Y^{ij}_{u}  \widehat{U}_i \widehat{Q}_j^a \widehat{H}_u^b 
              + Y^{ij}_e \widehat{E}_i \widehat{L}_j^b  \widehat{H}_d^a \right. \nonumber \\
              &+&\left. Y^{ij}_\nu 
\widehat{\nu}^c_i \widehat{L}^a_j \widehat{H}_u^b - \mu \widehat{H}_d^a \widehat{H}_u^b \right] 
+M_{R_i}\widehat{\nu}^c_i\widehat{X}_i+
\frac{1}{2}\mu_{X_i}\widehat{X}_i\widehat{X}_i  ~,
\label{eq:SuperPot}
\eea
\noindent where $i,j = 1,2,3$ denote generation indices.
In the above,  $\widehat H_d$ and $\widehat H_u$ are the down- and
up-type Higgs superfields, $\widehat L_i$ denotes the SU(2)
doublet lepton superfields.
$M_{R_i}$ represents the right-handed neutrino
mass term which conserves lepton number. 
Due to the presence of non-vanishing $\mu_{X_i}$, the total lepton number  $L$ is no longer a good
quantum number; nevertheless, notice that in our formulation
$(-1)^L$ is still a good symmetry.  
Without loss of generality, the terms $\widehat {\nu}^c_i
\widehat X_i$ and $\widehat X_i \widehat X_i$ are taken to be diagonal in generation space. 
Clearly, as $\mu_{X_i}\rightarrow 0$, lepton number conservation is restored,
since $M_R$ does not violate lepton number. 
Although in the present study we consider three generations of  $\widehat{\nu}^c$ and 
$\widehat{X}$, we recall that in the minimal version of the SUSY inverse seesaw (where only one generation of $\widehat{\nu}^c$ and 
$\widehat{X}$ is included), neutrino data can be accommodated~\cite{Hirsch:2009ra}. 

\noindent
The soft SUSY breaking Lagrangian can be written as
\beq
-{\mathcal L}_{\rm soft}=-{\mathcal L}^{\rm MSSM}_{\rm soft} 
         +  m^2_{\widetilde \nu^c} \widetilde\nu^{c\dagger}_i \widetilde\nu^c_i
         +m^2_X \widetilde X^{\dagger}_i \widetilde X_i
     + \left(A_{\nu}Y^{ij}_\nu \varepsilon_{ab}
                 \widetilde\nu^c_i \widetilde L^a_j H_u^b +
                B_{M_{R_i}} \widetilde\nu^c_i \widetilde X_i 
      +\frac{1}{2}B_{\mu_{X_i}}\widetilde X_i \widetilde X_i
      +{\rm h.c.}\right),
\label{eq:softSUSY}
\eeq
where ${\mathcal L}^{\rm MSSM}_{\rm soft}$ denotes the soft 
SUSY breaking terms of the MSSM. In the above, the singlet scalar states 
$\widetilde X_i$ and $\widetilde{\nu}^c_i$ are assumed to have flavour universal
masses, i.e.  
$m^2_{X_i}=m^2_{X}$ and $m^2_{\widetilde{\nu}^c_i}=m^2_{\widetilde{\nu}^c}$. The 
parameters $B_{M_{R_i}}$ and $B_{\mu_{X_i}}$ are the new terms involving the 
scalar partners of the sterile neutrino states (notice that while 
the former 
conserves lepton number, the latter gives rise to a  lepton number violating 
$\Delta L=2$ term). 
Working under the assumption of a flavour-blind mechanism for SUSY breaking, 
we will assume
universal boundary conditions\footnote{In our subsequent numerical  analysis, we will relax some of these universality conditions, considering non-universal soft breaking terms for the Higgs sector. In what concerns the right-handed sneutrino sector, we will assume that the corresponding soft-breaking masses hardly run between the GUT and the low-energy scale. } 
for the soft SUSY breaking parameters at 
some very high energy scale (e.g. the gauge coupling unification scale 
$\sim 10^{16}$ GeV),
 \beq
 m_\phi = m_0\,, M_\text{gaugino}= M_{1/2}\,, A_{i}= A_0\,.
 \eeq

Before addressing neutrino mass generation, a few comments on the nature
of the superpotential are in order. 
As can be seen from Eq.~(\ref{eq:SuperPot}), the two singlets 
$\widehat{\nu}^c_i$ and $\widehat X_i$ are differently treated in the sense that a $\Delta L = 2$
Majorana mass term is present for $\widehat X_i$ ($\mu_{X_i} \widehat X_i \widehat X_i$), while no 
$\mu_{\nu^c_i} \widehat{\nu}^c_i\widehat{\nu}^c_i$ is present in ${\mathcal W}$.
Although a generic superpotential with $(-1)^L$ should contain the latter term, let us notice that 
similar to what occurs for $\mu_{X_i}$, the absence of $\mu_{\nu^c_i}$ also enhances the symmetry 
of the model; moreover, we emphasise that it is the magnitude of
$\mu_{X_i}$ (and not that of $\mu_{\nu^c_i}$) which controls the size of the light
neutrino mass~\cite{Ma:2009gu,Bazzocchi:2010dt}. 
In view of this, and for 
the sake of simplicity, we have assumed $\mu_{\nu^c_i}= 0$ (considering non-vanishing, yet small values of 
$\mu_{\nu^c_i}$ would not change the qualitative features of the model). 
Although in our formulation we treat $\mu_{X_i}$ as
an effective parameter, its origin can be explained either dynamically or 
in a framework of SUSY Grand Unified Theories (GUT) 
\cite{Ma:2009gu,Bazzocchi:2010dt,Dev:2009aw}. Furthermore
$\mu_{\nu^c_i} \ll \mu_{X_i}$ can also be realised in extended frameworks~\cite{Ma:2009gu}.

\medskip
In order to illustrate the pattern of light neutrino masses in the 
inverse seesaw model and how it is related to the lepton number 
violating parameter $\mu_{X_i}$, we consider the one-generation case.
In the 
$\{\nu,{\nu^c},X\}$ basis
the $(3 \times 3)$ neutrino mass matrix can be written as
\begin{eqnarray}
{\cal M}&=&\left(
\begin{array}{ccc}
0 & m_D & 0 \\
m_D & 0 & M_R \\
0 & M_R & \mu_X \\
\end{array}\right) \ ,
\label{nmssm-matrix}
\end{eqnarray}
with $m_D= Y_\nu v_u$, yielding the mass eigenvalues ($m_1 \ll m_{2,3}$):
\begin{eqnarray}
 m_1 = \frac{m_{D}^2 \mu_X}{m_{D}^2+M_{R}^2} \, , ~~~~ 
 m_{2,3} = \mp \sqrt{M_{R}^2+ùm_{D}^2} + 
\frac{M_{R}^2 \mu_X}{2 (m_{D}^2+M_{R}^2)} \, . 
\label{masses}
\end{eqnarray}
The above equation clearly reveals that  the lightness of the
smallest eigenvalue $m_1$ is indeed due to the smallness of $\mu_X$ ($\mu_X\simeq m_1$). 
Thus the lepton number conserving mass parameters ($m_D$  and $M_R$) are 
completely unconstrained in this model.
{Finally, it is worth noticing  that the} effective right-handed sneutrino mass term (Dirac-like) is given by
$M^2_{\widetilde \nu^c_i} = m^2_{\widetilde \nu^c} + M_{R_i}^2 + 
\sum_j { |Y^{ij}_\nu|^2 v_u^2}$. 
Assuming $M_R \sim {\mathcal{O}}$(TeV), the effective mass term will 
not be very large, in clear contrast
to what occurs in the standard (type I) SUSY seesaw. 
In our analysis, we will be particularly interested in the r\^ole 
of such a light sneutrino (i.e.  
$M^2_{\widetilde \nu^c} \sim M^2_\text{SUSY}$) in the enhancement of Higgs 
mediated contributions to 
lepton flavour violating observables.

\section{Lepton flavour violation: Higgs-mediated contributions}
\label{lfv:diag}

In the SUSY seesaw framework, 
the only source of flavour violation is encoded in 
the neutrino Yukawa couplings (which are necessarily non-diagonal to 
account for neutrino oscillation data); even under the assumption of 
universal soft breaking terms at the GUT scale, radiative effects 
proportional to $Y_\nu$ induce flavour violation in the slepton mass 
matrices, which in turn give rise to slepton mediated cLFV 
observables~\cite{bm,hisano}. 
As an example, in the leading logarithmic approximation, the RGE 
corrections to the left-handed slepton soft-breaking masses are given by
\begin{eqnarray}
(\Delta m_{\widetilde{L}}^2)_{ij}&\simeq&
-\frac{1}{8\pi^2}(3m_0^2+A_0^2) 
(Y_\nu^\dagger L Y_\nu)_{ij} \,, ~~ L=\ln\frac{M_{GUT}}{M_{R}} \,
\nonumber \\ 
&=&\xi (Y^\dagger_\nu Y_\nu)_{ij}.
\label{slepmixing}
\end{eqnarray}  
(For simplicity, in the above we are implicitly assuming a 
degenerate right-handed neutrino spectrum, $M_{R_i}=M_{R}$.)
The RGE-induced flavour violating entries, $(\Delta m_{\widetilde{L}}^2)_{ij}$, 
give rise to the dominant contributions to low-energy flavour 
violating observables in the charged lepton sector, such as 
$\ell_i \to \ell_j \gamma$ (mediated by chargino-sneutrino and 
neutralino-slepton loops) 
and $\ell_i \to \ell_j \ell_k \ell_m$ (from photon, $Z$ and 
Higgs mediated penguin diagrams). 

Compared to the standard (type I) SUSY seesaw, where 
$M_{R}\sim10^{14}$ GeV, the inverse seesaw is characterised by a right
handed neutrino mass scale $M_{R}\sim\mathcal{O}(\text{TeV})$ and this 
in turn leads to an enhancement of the factor $\xi$, (see Eq.~(\ref{slepmixing})), and hence to all low-energy cLFV observables, 
in the latter framework. Furthermore, having right-handed sneutrinos 
whose mass is of the same order of  the other sfermions, 
i.e.  $M^2_{\widetilde \nu^c} \sim M^2_\text{SUSY}$, 
the $\widetilde \nu^c$-mediated processes are no longer suppressed, and 
might even significantly contribute to the low-energy flavour 
violating observables.
Here, we focus on the impact of  such a light $\widetilde \nu^c$ 
in the Higgs mediated processes which are expected to be  important 
in the large $\tan \beta$ regime.
 
Although at tree level Higgs-mediated neutral currents are flavour conserving, 
non-holomorphic Yukawa interactions
of the type $\bar D_RQ_LH_u^*$ can be induced at the one-loop level, as first 
noticed in~\cite{hrs}. 
In the large $\tan\beta$ regime,  in addition to providing significant corrections 
to the masses of the $b$-quark, 
these non-holomorphic couplings have an impact on $B^0-\bar B^0$ mixing and 
flavour violating  decays, in particular 
$B_s\rightarrow \mu^+\mu^-$ 
\cite{Choudhury:1998ze,babu:quark,isidori,buras,dedes}.
Similarly, in the lepton sector, the origin of the Higgs-mediated flavour violating couplings
can be traced to a non-holomorphic Yukawa term of the form 
$\bar E_RLH_u^*$~\cite{babu-kolda}. Other than the 
corrections to the 
$\tau$ lepton mass, these new couplings  give rise to additional 
contributions to several cLFV processes mediated by Higgs exchange. 
In particular $B_s \rightarrow \mu\tau$, $B_s \rightarrow e\tau$ (the so-called 
double penguin processes) were considered in~\cite{dedes}, while $\tau \rightarrow \mu \eta$ was 
studied in~\cite{sher}. 
A detailed analysis of the several $\mu-\tau$ lepton
flavour violating processes, namely $\tau \rightarrow \mu X$ 
($X = \gamma,e^+e^-,\mu^+\mu^-,\rho,\pi,\eta,\eta^\prime$) can be found in~\cite{rossi_anatomy}.

Even though the flavour violating processes in the quark and lepton sectors 
have a similar diagrammatic origin, the 
source of flavour violation is different in each case. 
In the quark sector, trilinear soft SUSY breaking couplings 
involving up-type squarks
provide the dominant source of flavour violation~\cite{babu:quark}, 
while in the lepton case, LFV stems
from the radiatively induced non-diagonal terms in the slepton masses 
(see Eq.~(\ref{slepmixing}))~\cite{babu-kolda}.

In the standard SUSY seesaw (type I), the 
term ${\widetilde \nu^c_{i}}H_u{\widetilde L_{Lj}}$ is usually neglected, as it is suppressed by the
very heavy right handed sneutrino masses (${M_{\widetilde \nu^c_{i}}} \sim 10^{14}$GeV).
However, in scenarios such as the inverse SUSY seesaw, where ${M_{\widetilde \nu^c_{i}}}\sim \mathcal{O}$(TeV), 
this term 
may provide the dominant contributions to Higgs mediated lepton 
flavour violation. 

\begin{figure}
\begin{center}
\begin{tabular}{cc}
\epsfig{file=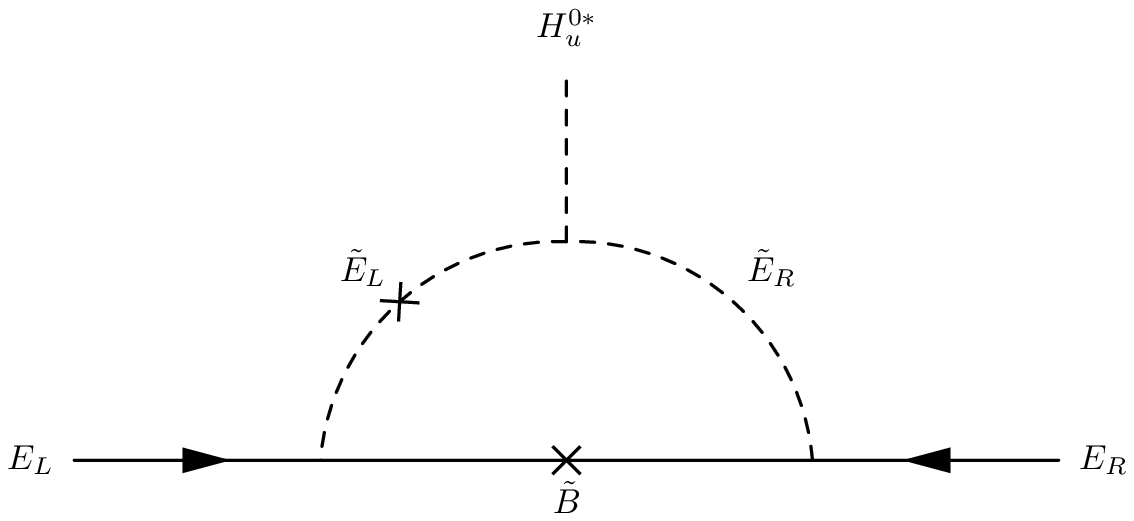, width=50mm, clip=} \hspace*{10mm}&\hspace*{10mm}
\epsfig{file=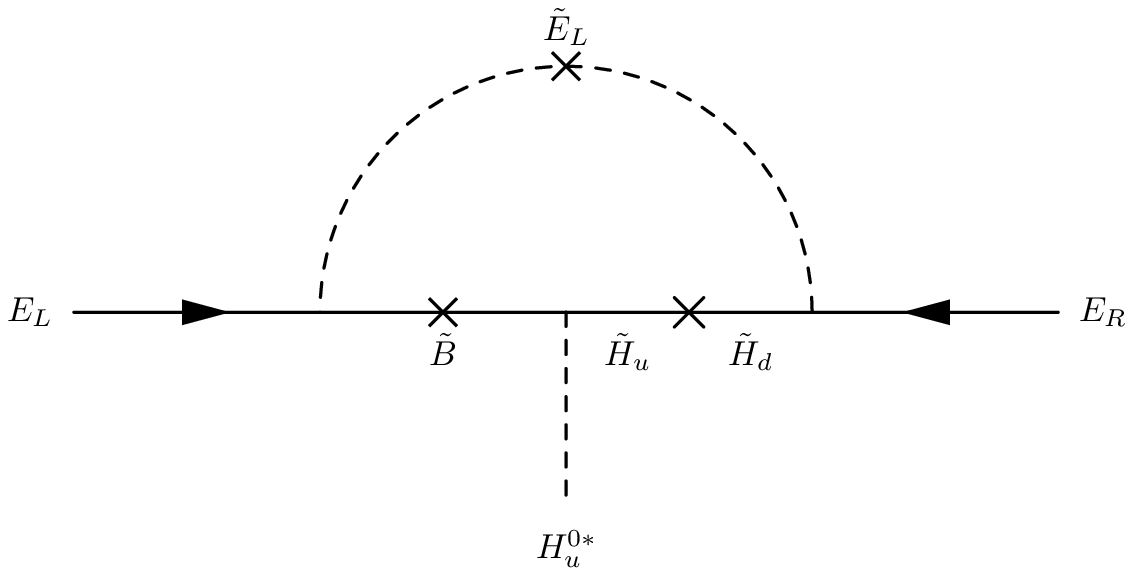, width=50mm, clip=}\vspace*{5mm} \\
\epsfig{file=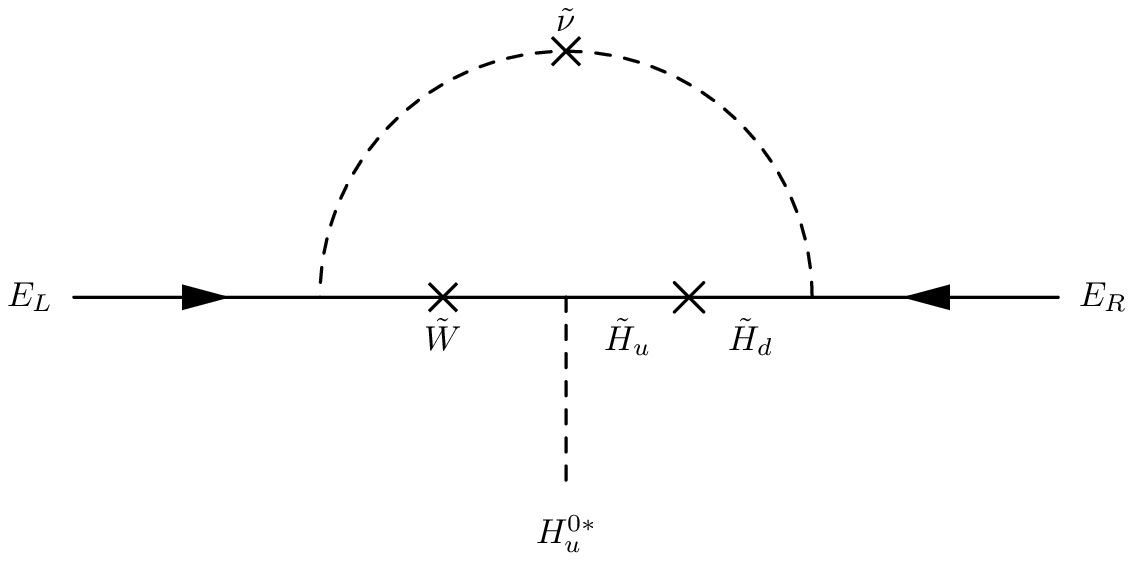, width=50mm, clip=} \hspace*{10mm}&\hspace*{10mm}
\epsfig{file=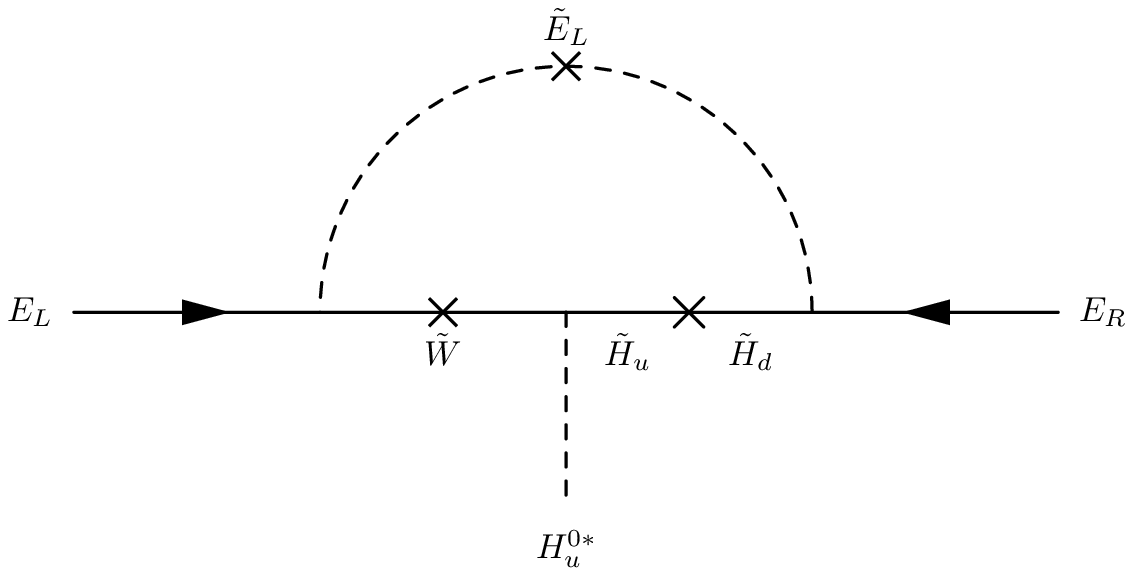, width=50mm, clip=} 
\end{tabular}
\end{center}
\caption{Diagrams contributing to $\epsilon_2$. Crosses 
on  scalar lines represent  LFV mass insertions $(\Delta m_{\widetilde{L}}^2)_{ij}$, while those on fermion lines denote chirality flips.}\label{1}
\end{figure}

The effective Lagrangian describing the couplings of the neutral Higgs fields
to the charged leptons is given by
\bea
-{\cal L}^\text{eff}=\bar E^i_R Y_{e}^{ii} \left[ 
\delta_{ij} H_d^0 + \left(\epsilon_1 \delta_{ij} + 
\epsilon_{2ij} (Y_\nu^\dagger Y_\nu)_{ij} \right) H_u^{0\ast }
\right] E^j_L + \text{h.c.}  \,. 
\label{Leff}
\eea
In the above,  the first term corresponds to the usual Yukawa interaction, 
while the coefficient $\epsilon_1$ encodes
the corrections to the charged lepton Yukawa couplings. In the basis 
where  the charged lepton Yukawa couplings are diagonal, 
the last term in Eq.~(\ref{Leff}), i.e. 
$\epsilon_{2ij} (Y_\nu^\dagger Y_\nu)_{ij}$, is in general  
non-diagonal, thus
providing  a new source of charged
lepton flavour violation through Higgs mediation. Its origin can be diagrammatically understood from Fig.\ref{1},
where flavour violation is parametrized  via a mass insertion
$(\Delta m_{\widetilde{L}}^2)_{ij}$ (see Eq. (\ref{slepmixing})).
The coefficient $\epsilon_2$ can be estimated as

\begin{figure}
\begin{center}
\epsfig{file=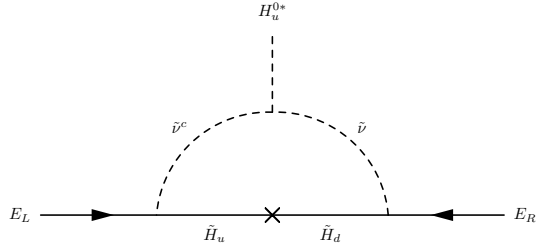, width=70mm, clip=}
\caption{Right-handed sneutrino contribution  to $\epsilon'_2$. 
This contribution is particularly relevant when 
$\widetilde \nu^c$ is light.}\label{2}
\end{center}
\end{figure}

\newpage
\bea
\epsilon_{2ij} &=& 
\frac{\alpha'}{8\pi} \xi  \mu M_1 
\left[
2F_2\left(M_1^2,m_{\widetilde E_{Lj}}^2,
m_{\widetilde E_{Li}}^2,m_{\widetilde E_{Ri}}^2\right)
-F_2 \left(\mu^2, m^2_{\widetilde E_{Lj}}, m^2_{\widetilde E_{Li}},M_1^2\right)
\right] + \cr
& &
\frac{\alpha_2}{8\pi}\xi  \mu M_2 
\left[
F_2\left(\mu^2,m^2_{\widetilde E_{Lj}}, m^2_{\widetilde E_{Li}}, M_2^2\right) +
2F_2\left(\mu^2,m_{\widetilde\nu_{Lj}}^2,m_{\widetilde \nu_{Li}}^2,M_2^2\right)
\right] \,,
\eea
where
\bea
F_2\left(x,y,z,w\right) =
-\frac{x \ln x}{(x-y)(x-z)(x-w)} -\frac{y\ln y}{(y-x)(y-z)(y-w)}+ 
(x\leftrightarrow z,
y\leftrightarrow w)\, . 
\eea
\noindent
Here, $M_1$ and $M_2$ are the masses of the electroweak gauginos at low energies.
On the other hand, the flavour conserving loop-induced form factor 
$\epsilon_{1}$ 
(notice that the diagrams of Fig.\ref{1} contribute to this form factor, 
but without  the slepton flavour mixings in the internal lines) 
can be expressed as~\cite{babu-kolda,dedes}
\bea
\epsilon_{1}&=&\frac{\alpha'}{8\pi}\mu M_1 \left[2 F_1\left(M_1^2,
m_{\widetilde E_L}^2,
m_{\widetilde E_R}^2\right) -
F_1\left(M_1^2,\mu^2,m^2_{\widetilde E_L}\right) + 2F_1\left(M_1^2,\mu^2,
m^2_{\widetilde E_R}\right)\right] \nonumber \\ 
&&+\frac{\alpha_2}{8\pi}\mu M_2\left[F_1\left(\mu^2,
m^2_{\widetilde E_L}, M_2^2\right) 
+ 2F_1\left(\mu^2,m_{\widetilde \nu_L}^2,M_2^2\right)\right],
\label{eq:eps1} 
\eea 
with
\bea
F_1\left(x,y,z\right)&=& 
-\frac{xy\ln (x/y)+yz\ln (y/z)+zx\ln (z/x)}
{(x-y)(y-z)(z-x)} \,.
\eea
\noindent
In the standard seesaw mechanism, the diagrams of 
Fig.~\ref{1} provide the only source for Higgs-mediated
lepton flavour violation. However, in the framework of the inverse SUSY seesaw,  
there is an additional diagram that 
may even account for the dominant Higgs-mediated
lepton flavour violation contribution: the sneutrino-chargino mediated loop, depicted in Fig.~\ref{2}. (Due to the large masses of $\widetilde \nu^c$ in the standard (type I) seesaw, 
this process provides negligible contributions, and is hence not taken into account.)

The effective Lagrangian terms encoding 
lepton flavour violation is accordingly modified as  
\bea
-{\cal L}^\text{LFV} = \bar E^i_R Y_{e}^{ii}\epsilon^\text{tot}_{2ij} (Y_\nu^\dagger Y_\nu)_{ij} H_u^{0\ast }
 E^j_L + \text{h.c.}  \,, 
\label{Leff1}
\eea
where $\epsilon^\text{tot}_2=\epsilon_2+\epsilon_2'$, $\epsilon_2'$ being  the 
contribution from the new diagram.
This contribution can be expressed as
\bea 
\epsilon'_{2ij}= \frac{1}{16\pi^2} \mu A_\nu 
F_1(\mu^2,m^2_{\widetilde \nu_i},M^2_{\widetilde \nu^c_j}).
\label{newdiag}
\eea
\noindent
In the above, we have parametrized the soft trilinear  term for the neutral leptons as $A_\nu Y_\nu$, where
$A_\nu$ is a flavour independent real mass term.

Below, we provide an
approximate estimate of the relative contributions
of the terms $\epsilon_{2}$ and $\epsilon'_{2}$:  for simplicity we take 
$M_{\widetilde \nu^c} \sim \mathcal{O}$(TeV) and assume common values 
for the masses of 
all SUSY particles and dimensionful terms $A_\nu$ at low energies, 
symbolically denoted by $A_\nu\sim \langle\widetilde m \rangle \sim M_\text{SUSY}$. 
In this limit,
the loop functions are given by 
$F_2\left(x,x,x,x\right) = \frac{1}{6x^2}$ and $F_1\left(x,x,x\right) = 
\frac{1}{2x}$. This leads to
\bea\label{eq:e2}
\epsilon_{2}= \frac{1}{8\pi} \xi {\widetilde m}^2 \left
(\frac{\alpha'}{6 {\widetilde m}^4}+3\frac{\alpha_2}{{6 \widetilde m}^4}\right)
\simeq -0.0007\,,
\eea
while
\bea
\epsilon'_{2}= \frac{1}{16\pi^2} {\widetilde m}^2 \frac{1}{2{\widetilde m}^2}
\simeq 0.003\,.
\eea
In this illustrative (leading order) calculation, we have 
assumed that at $M_\text{GUT}$, one has $A_0=0$, taking for 
the gauge couplings $\alpha_2=0.03$ and
$\alpha'=0.008$. 
Following  Eq.~(\ref{slepmixing}), and 
assuming $M_R=10^{3}~$GeV, one gets $\xi \sim -1.1 \,m_0^2$. Thus, at the leading 
order in the inverse seesaw,  the lepton flavour violation
coefficient becomes 
$|\epsilon^\text{tot}_{2}|= |\epsilon_{2}+\epsilon'_{2}| \simeq 2 \times 10^{-3}$.

For completeness, let us notice that in the standard seesaw model  (where sizable Yukawa couplings are typically associated to a right-handed neutrino mass scale $\sim10^{14}$ GeV), 
assuming the same amount of flavour violation as parametrized by $\xi$, one finds 
$|\epsilon^\text{tot}_{2}|= |\epsilon_{2}|\simeq 2 \times 10^{-4}$. 
This clearly reveals that in the inverse SUSY seesaw, $\epsilon^\text{tot}_{2}$ 
is enhanced by a factor of order $\sim 10$ compared to the standard seesaw.

The large enhancement of $\epsilon^\text{tot}_{2}$  
will have an impact regarding all 
Higgs-mediated lepton flavour violating observables. 
The computation of the cLFV observables requires specifying the 
couplings of the physical Higgs bosons to the leptons, in particular
$\bar E^i_{R}E^j_{L}H_k$ (where $H_k = h,H,A$). The effective 
Lagrangian describing this interaction can be derived from 
Eq.~(\ref{Leff}), and reads~\cite{babu-kolda,dedes} as
\bea
-{\cal L}^\text{eff}_{i\neq j} =
(2G_F^2)^{1/4} \,
\frac{m_{E_i} \kappa^E_{ij}}{\cos^2\beta}
\left(\bar E^i_{R}\,E^j_{L}\right)
\left[\cos(\alpha-\beta) h + \sin(\alpha-\beta) H - i A\right]+\text{h.c.}
\,,\,\,\,&&
\label{Leffl}
\eea
where $\alpha $ is the CP-even Higgs mixing angle and $\tan\beta=v_u/v_d$, and 
\bea
\kappa^E_{ij} &=& \frac{\epsilon^\text{tot}_{2ij} (Y^\dagger_\nu Y_\nu)_{ij}
}{
\left[1+\left(\epsilon_1+\epsilon^\text{tot}_{2ii}
(Y^\dagger_\nu Y_\nu)_{ii}\right)\tan\beta\right]^2 }\ \label{kappa}.
\eea

As clear from the above equation, large values of $\epsilon^\text{tot}_2$ lead to an augmentation 
of $\kappa^E_{ij}$. Given that the cLFV branching ratios are proportional to $({\kappa^E_{ij}})^2$,
a sizeable enhancement, as large as two orders of magnitude, is expected for all Higgs-mediated LFV observables.

\section{Higgs-mediated lepton flavour violating observables}\label{hmlfv}
Here we focus our attention on the cLFV observables where the dominant contribution to flavour 
violation arises from the 
Higgs penguin diagrams, in particular those involving $\tau$-leptons (due to the comparatively large value 
of  $Y_\tau$).

In what follows, we discuss some of these LFV decays in detail.
\begin{itemize}
\item
$\tau \rightarrow 3\mu$

\noindent
In the large $\tan\beta$ regime, 
Higgs-mediated flavour violating diagrams
would be particularly important in this decay mode. The branching ratio can be expressed as~\cite{babu-kolda,dedes} 
\bea
\text{Br}(\tau\to3\mu) &=&
\frac{G_F^2 \,m_\mu^2 \,m_\tau^7 \,\tau_\tau}{1536\,\pi^3 \cos^6\beta}\,
 |\kappa_{\tau\mu}^E |^2 
\left[\left(\frac{\sin(\alpha-\beta)\cos\alpha}{M_{H}^2} -
\frac{\cos(\alpha-\beta)\sin\alpha}{M_{h}^2}\right)^2 
+\frac{\sin^2\beta}{M_A^4}\right]
\cr
&\approx& 
\frac{G_F^2 \,m_\mu^2\, m_\tau^7\, \tau_\tau}{768\,\pi^3\, M_A^4}
 |\kappa_{\tau\mu}^E |^2 \tan^6\beta \,.
\eea
In the above,  $\tau_\tau$ is the $\tau$ life time and the approximate result has been obtained in the large $\tan\beta$ regime.  For other Higgs-mediated lepton flavour violating  3-body decays, 
$\tau \rightarrow e\mu\mu$,
$\tau \rightarrow 3e$ or 
$\mu \rightarrow 3e$, their corresponding branching ratios  can easily be obtained  with the appropriate
kinematic factors and the flavour changing factor $\kappa$. 
While $\text{Br}(\tau \rightarrow e\mu\mu)$ can be as large as  
$\text{Br}(\tau \rightarrow 3\mu)$ when $(Y^\dagger_\nu Y_\nu)_{13}\sim O(1)$ 
(which is possible in the case of  an inverted hierarchical light neutrino spectrum), 
other flavour 
violating decays  with final state electrons  such as $\mu \rightarrow 3e$ 
are 
considerably suppressed due to the smallness of the Yukawa couplings. 
\item
$B_s\to \ell_i \ell_j$

\noindent
$B$ mesons can also have  Higgs-mediated LFV decays, which are  significantly enhanced in the large $\tan\beta$ regime. The branching fraction is given by 
\bea
\text{Br}(B_s\to \ell_i \ell_j) &=& \frac{G_F^4 \,M^4_W}{8\,\pi^5}\,
|V_{tb}^*V_{ts}|^2\, M_{B_s}^5 \,f_{B_s}^2\, \tau_{B_s} \biggl (\frac{m_b}{m_b+
m_s}\biggr )^2 \nonumber \\[3mm] &\times & 
\sqrt{ \biggl [1-\frac{(m_{\ell_i}+m_{\ell_j})^2}{M_{B_s}^2}\biggr ]
\biggl [1-\frac{(m_{\ell_i}-m_{\ell_j})^2}{M_{B_s}^2}\biggr ] }       
\nonumber \\[3mm] &\times & 
\Biggl \{ \biggl (1-\frac{(m_{\ell_i}+m_{\ell_j})^2}{M_{B_s}^2}\biggr ) 
|c^{ij}_{S}|^2
+\biggl ( 1-\frac{(m_{\ell_i}-m_{\ell_j})^2}{M_{B_s}^2}\biggr ) |c_{P}^{ij}|^2
\Biggr \} \,,
\label{brll}
\eea
where $V_{ij}$ represents the Cabibbo-Kobayashi-Maskawa (CKM) matrix,  
$M_{B_s}$ and $\tau_{B_s}$ respectively denote the mass and lifetime of the 
$B_s$ meson, while $f_{B_s}=230\pm 30$ MeV~\cite{fbs} is the ${B_s}$ meson 
decay constant and $c_{P}^{ij}$, $c_{S}^{ij}$ are the form factors.
As an example, the lepton flavour violating (double-penguin) $B_s\to \mu\tau$ decay can be computed with the following form factors~\cite{dedes}: 
\bea
c_S^{\mu\tau}=c_P^{\mu\tau} &=&
\frac{\sqrt{2}\,\pi^2}{G_F \,M^2_W}
\frac{m_\tau \, \kappa_{bs}^d \,\kappa_{\tau\mu}^{E \ast}}
{\cos^4\beta \,\bar\lambda^t_{bs}}
\left[\frac{\sin^2(\alpha-\beta)}{M^2_{H}}+
\frac{\cos^2(\alpha-\beta)}{M^2_{h}} + \frac{1}{M^2_{A}} \right]
\nonumber \\[3mm]
&\approx&
\frac{8 \,\pi^2 \,m_\tau \, m_t^2}{ M^2_W}
\frac{ \epsilon_Y~ \kappa_{\tau\mu}^{E} ~\tan^4\beta }
{\left[1+(\epsilon_0+\epsilon_Y Y^2_{t} )\tan\beta\right]
\left[1+\epsilon_0 \tan\beta\right]}
\frac{1}{M^2_{A}} \,.
\label{ctm}
\eea
Here, $\kappa_{bs}^d$ represents the flavour mixing in the quark
sector while $\bar\lambda^t_{bs} = V^*_{tb}V_{ts}$. Similarly, 
$\epsilon_0$ and $\epsilon_Y$ are the down type quark form factors mediated
by gluino and squark exchange diagrams. 
The final result was, once again, derived in the large $\tan\beta$ regime.
The branching fractions of other flavour violating decays such as  
$\text{Br}(B_{d,s}\rightarrow \tau e)$, would receive identical contribution from the
Higgs penguins. 
Likewise, the  $\text{Br}(B_{d,s}\rightarrow \mu e)$  
can be calculated using the appropriate form factors and lepton masses;
as expected, these will be suppressed when compared to $\text{Br}(B_{d,s}\rightarrow \tau \mu)$.

\item
$\tau\to \mu P$

\noindent
Similar to what occurred in the previous processes, virtual Higgs exchange could also induce
decays such as $\tau \to \mu P$, where $P$ denotes a neutral pseudoscalar 
meson ($P=\pi, \eta, \eta')$. In the large $\tan\beta$ limit, where the  pseudoscalar Higgs couplings to down-type quarks are enhanced,
CP-odd Higgs boson exchanges provide the
dominant contribution to the $\tau \to \mu P$ decay. The coupling can be written as
\bea
-i (\sqrt{2} \,G_F)^{1/2}\tan\beta~ A(\xi_d \,m_d \,\bar d \,d +
\xi_s \, m_s \,\bar s \,s +\xi_b \,m_b \,\bar b \,b) + {\rm h.c.} .
\eea
Here, the parameters $\xi_d,\,\xi_s,\,\xi_b$ are  of order  $\mathcal{O}(1)$. 
Since we are mostly interested in the Higgs-mediated contributions, we estimate the amplitude of these processes in the limit when both $\tau \rightarrow 3 \mu$ and 
$\tau \to \mu P$ are indeed  dominated by the exchange of the scalar fields.
Accordingly, and following~\cite{rossi_anatomy}, one can write 
\bea
\frac{\text{Br}(\tau \to \mu \eta)}{\text{Br}(\tau \to 3 \mu )} 
&\simeq	&
36 \,\pi^2 \left(\frac{f^8_\eta \,m^2_\eta}{m_\mu\, m^2_\tau}\right)^2  
(1 - x_\eta)^2 
\left[\xi_s +\frac{\xi_b}{3}\left(1 +\sqrt2 \,\frac{f^0_\eta}{f^8_\eta}
\right)\right]^2 , \\
\label{taumu_eta}  
\frac{\text{Br}(\tau \to \mu \eta')}{\text{Br}(\tau \to \mu \eta)} 
& \simeq &	  
\frac2 9  \left( \frac{f^0_{\eta'}}{f^8_\eta}\right)^2 
\frac{m^4_{\eta'}}{m^4_\eta} \left(\frac{1 - x_{\eta'}}{1 - x_\eta} \right)^2
\left[\frac{
1 + \frac{3}{\sqrt2}\, \frac{f^8_{\eta'}}{f^0_{\eta'}} 
\left( \frac{\xi_s}{\xi_b}  +\frac13\right)}
{\frac{\xi_s}{\xi_b} + \frac13 + 
\frac{\sqrt2}{3} \,\frac{f^0_\eta}{f^8_\eta}}
\right]^2 , \\
\label{pieta}
\frac{\text{Br}(\tau\to \mu \pi)}{\text{Br}(\tau \to \mu \eta)} 
& \simeq & \frac4 3  \left(\frac{f_\pi}{f^8_\eta}\right)^2 \, 
 \frac{m^4_\pi}{m^4_\eta} ~   
(1 - x_\eta)^{-2}
\left[\frac{\frac{\xi_d}{\xi_b} \,\frac{1}{1+z} + \frac{1}{2}\,
(1 + \frac{\xi_s}{\xi_b}) 
\frac{1-z}{1+z}}
{\frac{\xi_s}{\xi_b} + \frac13 + 
\frac{\sqrt2}{3}\, \frac{f^0_\eta}{f^8_\eta}}
\right]^2 \,, 
\eea
where $z=m_u/m_d$, $m_\pi, \ f_\pi$ are the pion  mass and decay constant,  $m_{\eta,\eta'}$ 	are  the  masses of $\eta,\ \eta'$, $x_{\eta,\eta'}= m_{\eta,\eta'}^2/m_\pi^2$,  and 
$f^8_{\eta,\eta'}$ and $f^0_{\eta,\eta'}$ are evaluated from the corresponding matrix elements. 
As first discussed in~\cite{sher}, and taking $\xi_s,\xi_b \sim 1$ and fixing the other parameters as in~\cite{rossi_anatomy}, one finds  $\frac{\text{Br}(\tau \to \mu \eta)}{\text{Br}(\tau \to 3 \mu )} \simeq
5$. The other branching fractions such as 
$\text{Br}(\tau \to \mu \eta',\mu \pi)$  are considerably  suppressed compared to 
$\text{Br}(\tau \to \mu \eta)$. 
While the ratio $\frac{\text{Br}(\tau \to \mu \eta')}{\text{Br}(\tau \to \mu \eta)}$ 
can be as large as $6\times 10^{-3}$, 
$\frac{\text{Br}(\tau\to \mu \pi)}{\text{Br}(\tau \to \mu \eta)}$ 
would approximately lie in the range  
$10^{-3}-4\times 10^{-3}$~\cite{rossi_anatomy}. 
Since all these ratios 
are independent of $\kappa_{\tau\mu}^{E}$, the above quoted numbers can also be
applied to the present framework. 
However, an enhancement in the 
$\text{Br}(\tau \to 3 \mu)$, due to 
the large values of  $\kappa_{\tau\mu}^{E}$, 
would also imply sizeable values of $\text{Br}(\tau \to \mu \eta)$.

\item $H_k\to \mu\tau\;(H_k = h,H,A)$

\noindent
The branching ratios of flavour violating 
Higgs decays provide another interesting
probe of lepton flavour violation. 
Following~\cite{rossi_higgs},  the branching fraction $H_k\to \mu\tau$ 
(normalised
to the flavour conserving one  
$H_k\to \tau \tau$) can be cast as:
\bea
{\text{Br}(H_k\to \mu\tau)}
=  \tan^2\beta~ (|\kappa_{\tau\mu}^{E}|^2 ) 
 ~ C_\Phi~ { \text{Br}(H_k\to \tau\tau)}  \, ,
\eea
where we approximated $1/\cos^2\beta \simeq \tan^2\beta$. 
The coefficients $C_\Phi$  are given by:
\bea
C_h = \left[\frac{\cos(\beta - \alpha)}{\sin\alpha}\right]^2, ~~~~
C_H = \left[\frac{\sin(\beta - \alpha)}{\cos\alpha}\right]^2, ~~~~
C_A = 1.
\eea

\end{itemize}

\section{Results and Discussion}\label{sec:discussion}
As discussed in Section~\ref{lfv:diag}, in the inverse supersymmetric seesaw, Higgs-mediated contributions can lead to an enhancement of several LFV observables by as much as two orders of magnitude, compared to what is expected in the standard SUSY seesaw.

\begin{figure}
\begin{center}
\epsfig{file=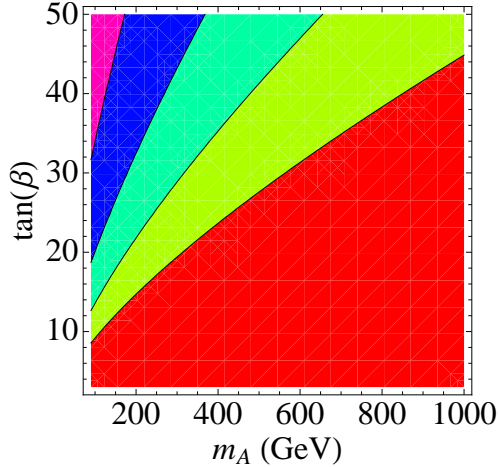, width=65mm}
\end{center}
\caption{Branching ratio of the process $\tau \rightarrow 3 \mu$ as a function of $m_A$ (GeV) and $\tan \beta$.
From left to right, the contours correspond to $\text{Br}(\tau \rightarrow 3 \mu) =2.1 \times 10^{-8}$, $10^{-9}$, $10^{-10}$, $10^{-11}$. The purple region has already been experimentally excluded\cite{Hayasaka:2010np}.}\label{3}
\end{figure}

As expected from the analytical study of Section~\ref{hmlfv}, 
$m_A$ and $\tan \beta$ are the most relevant parameters
in the Higgs-mediated flavour violating processes. To better illustrate this, 
in Fig.~\ref{3} we study the dependence of Br($\tau \rightarrow 3 \mu$) on the aforementioned parameters. 
We have assumed a common value for the squark masses, $m_{\widetilde q}\sim \text{TeV}$, 
while for left- and right-handed sleptons we take $m_{\widetilde \ell}\sim 400~\text{GeV}$ and $M_{\widetilde \nu^c}\sim 3~\text{TeV}$ for the right handed sneutrinos. 
The contours correspond to different values of the branching ratios (the purple region has already been experimentally excluded). From this figure one can easily identify the regimes for $m_A$ and $\tan\beta$ which are associated to values of the LFV observables within reach of the present and
future experiments. 

In what follows, we numerically evaluate some LFV observables.  Concerning the 
mSUGRA parameters 
(and instead of scanning over the parameter space), we have selected a 
few benchmark points~\cite{AbdusSalam:2011fc} that already take into account 
the most recent LHC constraints~\cite{cms_atlas}. 
We have also considered the case in which the GUT scale universality conditions are 
relaxed for the 
Higgs sector, i.e.  scenarios of Non-Universal Higgs Masses (NUHM), as this allows to 
explore the 
impact of a light CP-odd Higgs boson.
In Table~\ref{tab:sfp10.1}, we list the chosen points: CMSSM-A and CMSSM-B respectively 
correspond to 
the 10.2.2 and 40.1.1 benchmark points in~\cite{AbdusSalam:2011fc}, while NUHM-C is an example of a non-universal scenario.
\begin{table}[htb!]
\begin{center}
\begin{tabular}{|c||c||c|c||c|c|c|c|c|}
\hline
Point & $\tan\beta$ & $m_{1/2}$ & $m_0$ & $m^2_{H_U}$& $m^2_{H_D}$ &$A_0$  &$\mu$&$m_A$ \\
\hline
\hline
 CMSSM-A &10& 550 & 225 & $(225)^2$ &$(225)^2$ &0 &690 & 782 \\
\hline
\hline
CMSSM-B &40& 500 & 330 & $(330)^2$&$(330)^2$ &-500&698 &604 \\
\hline
NUHM-C &15& 550 & 225 &$(652)^2$ &$-(570)^2$ &0&478&150  \\
\hline
\hline
\end{tabular}
\caption{Benchmark points used in the numerical analysis (dimensionful parameters in GeV).
CMSSM-A and CMSSM-B correspond to 10.2.2 and 40.1.1 benchmark 
points of~\cite{AbdusSalam:2011fc}. \label{tab:sfp10.1} }
\end{center}
\end{table}

For each point considered, the low-energy SUSY parameters were obtained using SuSpect~\cite{suspect}.
 In what concerns the evolution of the soft-breaking right-handed sneutrino masses $m_{\tilde \nu^c}^2$, we
   have assumed that the latter hardly run between the GUT scale and the low-energy one. 
  The flavour-violating charged 
slepton parameters (e.g. $(\Delta m_{\widetilde{L}}^2)_{ij}$ or $\xi$), were estimated at the 
leading order using Eq.~(\ref{slepmixing}). Concerning NUHM, we use the same value of $\xi$ as for CMSSM-A. Here, we
are particularly interested to study the effect of light CP-odd Higgs boson and this naive
approximation will serve our purpose. Furthermore, we use the mass insertion approximation, 
assuming that mixing between left and right chiral 
slepton states are relatively small. In computing the branching fractions and the 
flavour violating factor $\kappa^E_{ij}$ 
we have assumed (physical) right-handed sneutrino masses
$M_{\widetilde \nu^c} \approx 3$ TeV and $\left(Y_\nu^{\dagger}Y_\nu \right) = 0.7$, in agreement with low-energy neutrino data as well as other low-energy  constraints, which are particularly relevant in the inverse seesaw case such as Non-Standard Neutrino Interactions bounds \cite{NSI}. Moreover, in our numerical analysis, we have 
fixed the trilinear soft breaking parameter $A_\nu = - 500$ GeV (at the SUSY scale).

We now proceed to present our results for the flavour violating observables
discussed in Section~\ref{hmlfv}. 
In Table~\ref{4}, we collect the values of the different branching ratios, as obtained for the considered benchmark points of Table~\ref{1}.  We have also presented the corresponding current experimental bounds and future sensitivity.

\begin{table}[htb]
\begin{center}
 \begin{tabular}{|c|c|c|c|c|c|}
  \hline
    LFV Process & Present Bound & Future Sensitivity & CMSSM-A & CMSSM-B & NUHM-C\\
  \hline
    $\tau \rightarrow \mu \mu \mu$ & $2.1\times10^{-8}$\cite{Hayasaka:2010np} & $8.2 \times 10^{-10}$ \cite{OLeary2010af} & $1.4 \times 10^{-15}$ & $3.9 \times 10^{-11}$ & $8.0 \times 10^{-12}$ \\
    $\tau^- \rightarrow e^- \mu^+ \mu^-$ &  $2.7\times10^{-8}$\cite{Hayasaka:2010np} & $\sim 10^{-10}$ \cite{OLeary2010af} & $1.4 \times 10^{-15}$ & $3.4 \times 10^{-11}$ & $8.0 \times 10^{-12}$ \\
    $\tau \rightarrow e e e$ & $2.7\times10^{-8}$\cite{Hayasaka:2010np} &  $2.3 \times 10^{-10}$ \cite{OLeary2010af} & $3.2 \times 10^{-20}$ & $9.2 \times 10^{-16}$ & $1.9 \times 10^{-16}$ \\
    $\mu \rightarrow e e e$ &  $1.0 \times 10^{-12}$\cite{Bellgardt:1987du} &  & $6.3 \times 10^{-22}$ & $1.5 \times 10^{-17}$ & $3.7 \times 10^{-18}$ \\
    $\tau \rightarrow \mu \eta$ & $2.3\times 10^{-8}$\cite{arXiv:1011.6474} & $\sim 10^{-10}$ \cite{OLeary2010af} & $8.0 \times 10^{-15}$ & $3.3 \times 10^{-10}$ & $4.6 \times 10^{-11}$ \\
    $\tau \rightarrow \mu \eta^\prime$ & $3.8\times 10^{-8}$\cite{arXiv:1011.6474} & $\sim 10^{-10}$ \cite{OLeary2010af} & $4.3 \times 10^{-16}$ & $1.1\times 10^{-10}$ & $3.1 \times 10^{-12}$ \\
    $\tau \rightarrow \mu \pi^{0}$ & $2.2\times 10^{-8}$\cite{arXiv:1011.6474} & $\sim 10^{-10}$ \cite{OLeary2010af} & $1.8 \times 10^{-17}$ & $8.5 \times 10^{-13}$ & $1.0 \times 10^{-13}$ \\
    $B^{0}_{d} \rightarrow \mu \tau$ & $2.2\times 10^{-5}$\cite{arXiv:0801.0697} & & $2.7 \times 10^{-15}$ & $8.5 \times 10^{-10}$ & $2.7 \times 10^{-11}$ \\
    $B^{0}_{d} \rightarrow e \mu$ & $6.4\times 10^{-8}$\cite{arXiv:0901.3803} & $1.6\times 10^{-8}$\cite{CERN-LHCB-2007-028} & $1.2 \times 10^{-17}$ & $3.1 \times 10^{-12}$ & $1.2 \times 10^{-13}$ \\
    $B^{0}_{s} \rightarrow  \mu \tau$ & & & $7.7 \times 10^{-14}$ & $2.5 \times 10^{-8}$ & $7.8 \times 10^{-10}$ \\
    $B^{0}_{s} \rightarrow e \mu$ & $2.0\times 10^{-7}$\cite{arXiv:0901.3803} & $6.5\times 10^{-8}$\cite{CERN-LHCB-2007-028} & $3.4 \times 10^{-16}$ & $8.9 \times 10^{-11}$ & $3.4 \times 10^{-12}$ \\
    $h \rightarrow \mu \tau$ & & & $1.3 \times 10^{-8}$ & $2.6 \times 10^{-7}$ & $2.3 \times 10^{-6}$\\
    $A,H \rightarrow \mu \tau$ & & & $3.4 \times 10^{-6}$ & $1.3 \times 10^{-4}$ & $5.0 \times 10^{-6}$\\
  \hline
 \end{tabular}
\end{center}
  \caption{Higgs-mediated contributions to the branching ratios of several lepton flavour violating 
processes, for the different benchmark points of Table~\ref{1}. We also present the current experimental 
bounds and future sensitivities for the LFV observables.}\label{4}
\end{table}

From Table~\ref{4}, one can verify that from an experimental point of view, 
the most promising channel in the supersymmetric inverse seesaw is $\tau \rightarrow \mu \eta$ 
which could be tested at the next generation of $B$ factories. 
The $B^{0}_{d,s} \rightarrow  \mu \tau$ decay might also be interesting, but little conclusions can be drawn due to lack of information concerning the future sensitivities.

It is important to stress that the numerical results summarised in  Table~\ref{4} correspond to considering 
{\it only} Higgs-mediated contributions. In the low $\tan \beta$ regime,
photon- and $Z$-penguin diagrams may induce comparable or even larger contributions to the observables, and potentially enhance the branching fractions. Thus, the results
for small $\tan \beta$ should be interpreted as
conservative estimates,  representing only partial
contributions. 
For large $\tan\beta$ values, Higgs penguins do indeed provide the leading contributions.
Comparing our results with those obtained for a
type I SUSY seesaw at high scales (or even with a TeV scale SUSY seesaw), 
we find a large enhancement
of the branching fractions in the inverse seesaw framework.

Another interesting property of the Higgs-mediated processes is that the corresponding 
amplitude strongly depends on 
the chirality of the heaviest lepton (be it the decaying lepton, or the heaviest 
lepton produced in $B$ decays). 
Considering the decays of a left-handed lepton $\ell^i_{L} \rightarrow \ell^j_{R} X$, one finds that 
the corresponding branching ratios would be suppressed by a factor $\frac{m_{\ell^j}^2}{m_{\ell^i}^2}$ compared to those of the right-handed lepton $\ell^i_{R} \rightarrow \ell^j_{L} X$. 
This can induce an asymmetry that potentially allows to identify if 
 Higgs mediation is the dominant contribution to the LFV observables. Furthermore
this asymmetry would be 
more pronounced in the inverse-seesaw framework.

Due to its strong enhancement of the Higgs-penguin contributions, if realised in Nature, 
the inverse seesaw offers a unique framework to test Higgs effects in LFV processes. 
In fact, and as discussed in~ \cite{hisano}, if photon penguins provide the dominant
contribution to both $\text{Br}(\tau \rightarrow 3 \mu)$ and $\text{Br}(\tau \rightarrow 
\mu \gamma)$, then the latter observables are strongly correlated,  
$\frac {\text{Br}(\tau \rightarrow 3 \mu)} {\text{Br}(\tau \rightarrow 
\mu \gamma)} \sim 0.003$ (see \cite{hisano}). On the other hand, if the dominant contribution to 
the three-body decays arises from Higgs penguins, the correlation no longer holds, and the latter ratio can be significantly enhanced. This would be the case of the present framework.

\section{Conclusions}\label{sec:conclusions} 
If observed, charged lepton flavour violation  clearly signals the presence of new physics. 
In this work, we have studied Higgs-mediated LFV processes  in the framework of the supersymmetric inverse seesaw. TeV scale right-handed neutrinos (and hence light right-handed sneutrinos) offer the possibility to enhance the Higgs-mediated contributions to several LFV processes. As shown in this work, in the inverse SUSY seesaw, LFV branching ratios can be enhanced by as much as two orders of magnitude when compared to the standard (type I) SUSY seesaw.

\subsection*{Acknowledgements}
The authors are thankful to A. Vicente for many enlightening discussions. 
D.D. acknowledges financial support from the CNRS.  
This work has been  partly done
under the ANR project CPV-LFV-LHC NT09-508531.

\end{document}